\documentclass{aastex}
\usepackage[onecolumn]{emulateapj5}

\def\ltsim{\mathrel{\hbox{\rlap{\hbox{\lower4pt\hbox{$\sim$}}}\hbox{$<$}}}}
\def\gtrsim{\mathrel{\hbox{\rlap{\hbox{\lower4pt\hbox{$\sim$}}}\hbox{$>$}}}}
\def\kms{km~s$^{-1}$}

\submitted{Accepted for publication in ApJ, 18 July 2001}

\righthead{Warm Absorbers in AGN}
\lefthead{Krolik \& Kriss}

\begin{document}

\title{Warm Absorbers in AGN: A Multi-Temperature Wind}

\author{Julian H. Krolik}
\affil{Department of Physics and Astronomy, Johns Hopkins
University, Baltimore, MD 21218}
\email{jhk@pha.jhu.edu}

\author{Gerard A. Kriss\altaffilmark{1}}
\affil{Space Telescope Science Institute, 3700 San Martin Drive,
Baltimore, MD 21218}
\email{gak@stsci.edu}

\altaffiltext{1}{also, Department of Physics and Astronomy, Johns Hopkins
University, Baltimore, MD 21218}

\begin{abstract}

   Although soft X-ray absorption features in AGN were discovered almost
ten years ago, the nature and location of the gas creating them has remained
controversial.  Guided by the results of recent high-resolution X-ray
spectroscopy, we argue that these features are created in a multi-temperature
wind whose source of matter is photoionized evaporation from the inner
edge of the obscuring torus often found surrounding AGN.  Photoionized
evaporation in the presence of a copious mass source locks the ratio
of ionizing intensity to pressure to a critical value.  However,
a broad range of temperatures can all coexist in equilibrium for
this value of the ratio of ionizing intensity to pressure.  Consequently,
the flow is expected to be strongly inhomogeneous in temperature.
The inferred distance of this material from the source of ionizing
radiation depends on how much matter exists at the highest-obtainable
temperature.  This distance can be measured by monitoring how ionic
column densities respond to changes in the ionizing continuum on
timescales of days to years.
\end{abstract}

\section{Introduction}

X-ray spectra of active galactic nuclei (AGN) commonly reveal intrinsic
absorption by highly ionized or ``warm" gas.  First seen in ROSAT and
ASCA spectra \cite{Turner93,Reynolds95,Reynolds97,George98}, absorption edges of
\ion{O}{7} and \ion{O}{8} can be found in half or more of all type~1
Seyfert galaxies, and in a few quasars \cite{Mathur94}. 

     The location of this gas is still unknown.
%For typical ionization parameters of $U_X \sim 0.1$ (\cite{George98}),
%and gas densities of $10^4 - 10^{10}~\rm cm^{-3}$, the absorber is located
%anywhere from the broad-line region (BLR) to within or beyond the
%narrow-line region (NLR).
Suggestions range all the way from the broad-line
region (BLR), possibly as close as $\sim 0.01$~pc from the nucleus,
to distances of 10~pc or more.
Rapid apparent column density variations have been interpreted
as suggesting locations in the BLR \cite{Reynolds95,George98}.  On
the other hand, the same data also indicate that at least some of
the gas may be much farther away \cite{Otani96}.
Others have placed the gas at $\sim 1$~pc in the scattering
region posited by Seyfert galaxy unification schemes \cite{KK95},
or a factor of 10 farther away \cite{BKS00}.  Morales et al. (2000)
suggested that the absorber is spread all the way from
$< 0.05$~pc to $> 1$~pc.
Still less is known about the origin, dynamics,
or destiny of the gas producing warm absorber features.  Some of
the diverse suggestions include:
evaporation off ``bloated stars" in the BLR \cite{Netzer96};
gas, evaporated off the torus obscuring the nucleus, that becomes the
scattering gas seen
%the gas producing polarized reflected light
in type~2 Seyfert galaxies \cite{KK95};
and a wind driven off the accretion disk \cite{Elvis00,BKS00}.

    With the launch of {\it Chandra} and {\it XMM-Newton}, we now have
access to data of far better quality, and may reasonably hope to settle
some of these outstanding questions.  Most notably, the high resolution
available with the grating spectrographs on these spacecraft has permitted
detection of resonance line features
\cite{Kaastra00,Kaspi00,Kaspi01,Lee01,Coll01}.
As we will show later, measurement (or even bounds) on line widths and
shifts (i.e., relative to the host galaxy rest-frame) can provide
powerful hints toward answering many of the questions just posed.
It is our object here to show how this new data points
us toward a revision of the ``scattering wind" model.  Because this
model makes specific predictions about the character of variability
on measurable timescales, it should be easily testable in the
near future.
 
\section{The Model: A Critical-Pressure Wind}

\subsection{Phenomenological guidance}

    Measured column densities of different ionization stages seen in
absorption can be used to constrain photoionization properties of the
absorbing gas under the assumption that the species involved are in
ionization (and optionally also thermal) balance.  The relative abundances
of different species fix the ionization parameter $\xi \equiv L_{ion}/n r^2$;
the strengths of
the lines and edges fix the hydrogen column density $N$.  Here the
ionizing luminosity $L_{ion}$ is conventionally taken to be the
luminosity between 1 and 1000~Ryd, $n$ is the hydrogen volume density,
and $r$ is the distance from the nucleus to the photoionized gas.
Other, closely-related, versions of the ionization parameter are also used,
e.g., $U_{oxygen}~\equiv~\int_{0.538}^{10}\,
d\epsilon (L_\epsilon/ 4\pi r^2 \epsilon n c)$ \cite{George00},
where photon energy $\epsilon$ is measured in keV and $L_\epsilon$ is
the luminosity
per unit energy.  Its precise relation to $\xi$ depends on the shape of
the continuum spectrum.

   These quantities can then be used \cite{Turner93} to estimate the
fractional radial thickness occupied by absorbing gas
\begin{equation}
\label{fracrad}
\Delta r/r = \xi N r/L_{ion} .
\end{equation}
Placing the gas close to the nucleus implies a highly ``clumped"
structure, whereas larger distances imply a more volume-filling configuration.
The gas cannot, in any case, be farther than $r_{max} = L_{ion}/(N\xi)$
because $\Delta r/r \leq 1$.  Simple photoionization-fitting cannot
distinguish between a clumpy and a smooth configuration; it can only
define the link between the degree of clumpiness (or smoothness)
and the distance from the nucleus to the absorbing gas.

    The most commonly observed features (especially in ASCA data)
are those associated with \ion{O}{7} and \ion{O}{8}
\cite{Reynolds97, George98}.
In order to make them reasonably abundant, $\xi \sim 10$ -- $100$.
For this reason, many photoionization models of warm absorbers have
suggested values of $\xi$ in this range, e.g., \cite{Brandt97,
Reynolds97,George98,Mathur97}.  On the other hand, more recent work
has detected many other species of higher ionization level \cite{Coll01},
suggesting that there may be substantial amounts of gas with
$\xi \sim 100$ -- 1000 \cite{Kaastra00,Kaspi01}.
The maximum distance of the absorbing region in a Seyfert galaxy is then
$r_{max} \sim 30 L_{ion,44} N_{22}^{-1} \xi_{100}^{-1}$~pc, where the fiducial
quantities ($10^{44}$~erg~s$^{-1}$ for $L_{ion}$, $10^{22}$~cm$^{-2}$
for $N$, and 100 for $\xi$) have been chosen to be representative of
those commonly inferred.

    Line profiles convey additional information.  For example, Kaspi
et al. (2001) find that the absorption lines they detect have a mean
blueshift relative to the galaxy's systemic velocity of 610~\kms.  The
gas must, therefore, possess a net flow toward us.  The associated mass
outflow rate is
\begin{equation}
\label{massloss}
\dot M_{out} = 4\pi C \mu_H N r v = 0.72 C N_{22} r_{\rm pc} (v/500\hbox{~\kms}) 
             \hbox{~$M_{\odot}$~yr$^{-1}$},
\end{equation}
where its covering fraction around the nucleus is $C$, $\mu_H$ is the
mean mass per H atom, $r_{\rm pc}$ is the radius normalized to 1~pc,
and $v$ is the mean outflow speed.
Interestingly, the ratio of the outflow rate to the accretion
rate required to power the nuclear luminosity can be $\sim 100$:
\begin{equation}
\label{massratio}
{\dot M_{out} \over \dot M_{acc}} = 600 C v_{500}
\left({\eta_{acc} \over 0.1}\right)
\left({r \over r_{max}}\right)\left({L_{ion}/L_{bol} \over 0.5}\right)
\left({\xi \over 100}\right)^{-1},
\end{equation}
where $L_{bol}$ is the bolometric luminosity and $\eta_{acc}$ is the accretion
efficiency (cf. Krolik \& Begelman 1986, Reynolds 1997).

     The widths of the absorption lines appear to be similar to their
blueshifts; many absorption lines also have emission components to the
red (i.e., ``P Cygni"-type profiles) \cite{Kaspi01}.
The fact that the emission components are centered at the systemic velocity
and have widths comparable to the blueshifts is most simply understood as
indicating that emission and absorption are made by gas with similar
properties, but that the line emission sources are located to the side of
and behind the continuum source.

    All measures of gas velocity are roughly an
order of magnitude smaller than the typical speeds of gas in the
BLR.  Following the common rule-of-thumb that speeds diminish at
increasing distance from the central black hole, this contrast would
suggest that the absorbing gas is rather more distant than the BLR
\cite{Kaspi00}.  Still more speculatively, if velocities scale
$\propto r^{-1/2}$, the warm absorber material must be $\sim 100$
times farther from the nucleus than the BLR, i.e., in NGC~3783 it
must be parsecs from the nucleus if we adopt the reverberation-mapping
estimate of its BLR scale \cite{Reichert94}.

     On the basis of this evidence (modest velocity ouflow, maximum
distance $\sim 30 L_{ion,44}$~pc, moderately high ionization) and the
comparisons to observations presented in \S 3, we find the most
natural model to be a close relation to the one proposed by
Krolik \& Kriss (1995).  Again, we identify the warm absorber
with the hot scattering gas of Seyfert galaxy unification models.
However, as discussed in the next subsection, our view of its
thermodynamic state is somewhat different from the one expressed
six years ago.

\subsection{Thermodynamics of the warm absorber}

    In this model, the origin of the warm absorber gas is material
evaporated off the inner edge of the toroidal obscuration.
The position of this inner edge is determined by a combination of
dust sublimation and photoionization \cite{PV95}, but is expected
to be $\sim 1 L_{ion,44}^{1/2}$~pc.  The existence of rings of H$_2$O
masers at approximately this distance \cite{Greenhill96} is in
keeping with this expectation.  Thus, the source for the warm
absorbing gas would be roughly this far from the nucleus.

    The thermodynamics of this evaporating gas are best described by
an ionization parameter framed in terms of the gas pressure, such
as $\Xi \equiv L_{ion}/(4\pi r^2 c n kT)$ (Krolik, McKee \& Tarter 1981).
When $\Xi$
exceeds a critical value ($\Xi_c \sim 10$ -- 30 for typical AGN spectra),
the gas is no longer capable of maintaining a cool ($\sim 10^4$~K)
equilibrium temperature; instead its equilibrium temperature
rises very rapidly with increasing $\Xi$ until it approaches
the Compton temperature, which in AGN is generally several orders of
magnitude greater.

    The detailed character of this rise depends on the specific shape
of the continuum (and also on details of atomic physics that are imperfectly
known), but its qualitative behavior---a nearly vertical rise from
temperatures $\sim 3 \times 10^4$~K to $\sim 10^6$~K at
$\Xi \simeq \Xi_c$---is the same for almost any AGN-like spectrum.  In
Figure~1 we show $T(\Xi)$
(as computed using Version 2.1 of XSTAR, Kallman 2000) for two
continuum shapes based on data from the type~1 Seyfert galaxy NGC~3783, one
suggested by Kaiser et al. (2001), the other by Kaspi et al. (2001).
The rise is very similar for both spectra; almost the only difference
is the offset in $\Xi_c$.  The fiducial $\xi = 100$ corresponds to a
temperature part way up this rise, $1.5 \times 10^5$~K for the Kaiser
et al. spectrum, $5.5 \times 10^4$~K for the Kaspi et al. spectrum.
In terms of $\xi$, the vertical portion of the equilibrium curve runs
from $\xi \simeq 10$ to $\xi \simeq 1000$ in the case of the Kaiser
et al. spectrum, but from $\xi \simeq 40$ to $\xi \simeq 1500$ for
the Kaspi et al. spectrum.

\begin{figure*}[t]
\plottwo{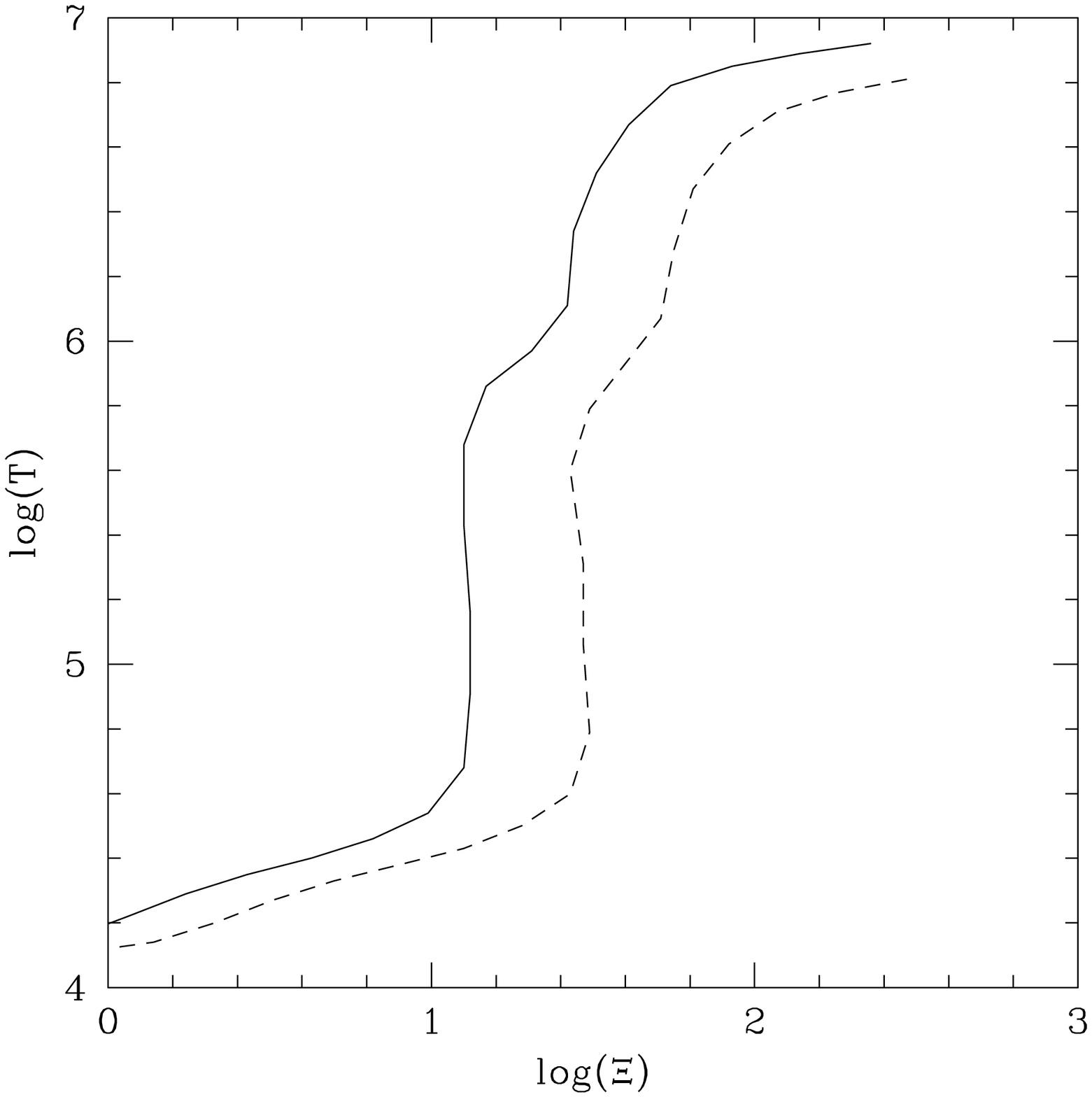}{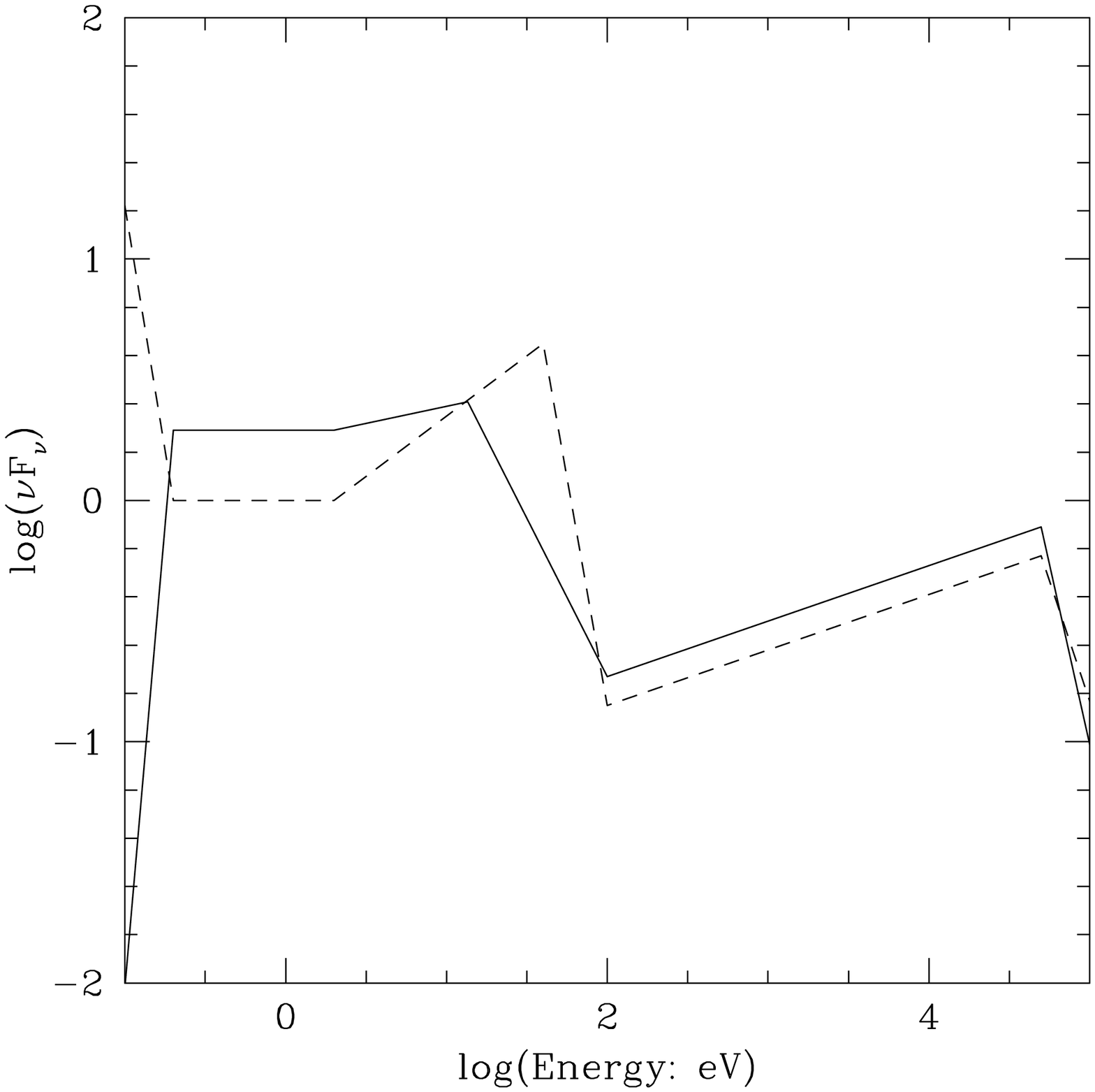}
\hbox{\hfill}{
\parbox{0.25in}{\hfill}
\parbox{3.25in}{
Fig.~1.---The equilibrium temperature as a function of $\Xi$
for the Kaiser et al. (2001) spectrum of NGC~3783 (solid curve) and the
Kaspi et al. (2001) spectrum (dashed curve).
%\phantom{Kaspi et al. (2001).  Both are plotted
%in terms of flux per logarithmic frequency interval, the
%vertical scale is arbitrary.}
}
\hfill
\parbox{3.25in}{Fig.~2---Proposed continuum shapes for NGC 3783: solid curve
by Kaiser et al. (2001), dashed curve by Kaspi et al. (2001).  Both are plotted
in terms of flux per logarithmic frequency interval, $\nu F_\nu$; the
vertical scale is arbitrary.}
\parbox{0.35in}{\hfill}
}
\end{figure*}

    In this paper we are primarily interested in material near and outside
the obscuring torus.  In keeping with theoretical expectations that the torus
should reradiate a substantial portion of the nuclear luminosity in
the infrared, we have required that in each spectrum the luminosity
at energies below 2~eV match the ionizing luminosity.  No alteration
in the Kaiser et al. spectrum was necessary to accomplish this, but to
achieve this with the Kaspi et al. spectrum we added a new component
at 0.1~eV.   Both spectra are shown in Figure~2.  The
additional infrared intensity makes little difference to the
equilibrium curve at temperatures below $\sim 10^6$~K; the primary
effect is to reduce the Compton temperature and create the ``jog"
toward higher $\Xi$ in the range 1 -- $3 \times 10^6$~K.
Closer to the nucleus, the infrared intensity would be relatively weaker,
leading to a higher Compton temperature and a longer section of
nearly-vertical temperature rise.
Throughout the remainder of this paper, specific numbers derived from
photoionization calculations assume the Kaiser et al. spectral shape.
%As a result, in the
%BLR the minimum $\Xi$ for which Compton equilibrium can exist should
%decrease, forcing the rising section of the equilibrium curve
%to ``double back" farther.

%\begin{figure*}
%\epsscale{0.70}
%\plotone{specplot.ps}
%\caption{Proposed continuum shapes for NGC 3783: solid curve
%by Kaiser et al. (2001), dashed curve by Kaspi et al. (2001).  Both are plotted
%in terms of flux per logarithmic frequency interval, $\nu F_\nu$; the
%vertical scale is arbitrary.\label{specplot}}
%\end{figure*}

Places where the slope $dT/d\Xi > 0$ are stable against isobaric thermal
perturbations, but places where the curve ``doubles back"
(i.e., where $dT/d\Xi < 0$) are thermally unstable.  As shown by the curves of
Figure~1, near the obscuring torus there is a significant stretch of the
equilibrium curve where the sign of $dT/d\Xi$ is ill-defined because
the temperature rises almost exactly vertically:
$d\Xi /dT \simeq 0$ for $10^{4.5}$~K$ \ltsim T \ltsim 10^{5.9}$~K (and
$|d\Xi/dT|$ remains very small up to $T \sim 10^{6.5}$~K).
This region is therefore better described as having approximately
marginal stability.
That is, thermal fluctuations neither grow nor decay exponentially;
there is little feedback in either sense.  Although the equilibrium
curves are never exactly vertical locally, because the overall sense
is very nearly vertical, tiny pressure perturbations in regions that
are nominally stable can nonetheless lead to large temperature changes.
Likewise, temperature perturbations in regions of nominal instability do not
lead to arbitrarily large temperature changes---at constant pressure,
material rises or falls along a line of constant $\Xi$, and cannot go far before
intersecting the equilibrium curve again, this time on a locally stable
segment.  It is in this sense that we describe this
portion of the equilibrium curve as being ``marginally stable".

    The reason to pay such close attention to the character of equilibria
with $\Xi \simeq \Xi_c$ is that when there is a copious source of gas
available for rapid heating (as the obscuring torus provides in this model),
rather than being a special value, $\Xi_c$ is a preferred value of the
ionization parameter.  Suppose that the pressure in the gas at the illuminated
face of the torus were very low, so that $\Xi > \Xi_c$.  Then this gas would
be subject to strong net heating.  As its temperature rises, it expands and
its opacity drops sharply.  Deep inside the torus, the large continuum optical
depth protects the gas and keeps it relatively cool (the most likely
temperature is $\ltsim 10^3$~K: Krolik \& Lepp 1989, Neufeld et al. 1994),
but as the opacity of matter at the illuminated edge is eliminated, new
gas is exposed and becomes available for ``evaporation".  As more and more
gas is removed from the torus and added to the hot phase, the pressure of
the hot phase rises because (as we shall show shortly), its heating time
is much shorter than the flow time.  Eventually, the pressure rises high
enough that $\Xi$ falls to $\Xi_c$.  Conversely, if the ambient pressure
were so high that $\Xi < \Xi_c$, gas outside the torus would cool and
condense, diminishing the pressure.  

    Thus, where the gas is injected, the ionization parameter is fixed
at $\Xi_c$.  Moreover, with one relatively modest proviso, it remains
at that value even as the gas flows away from its injection point.  So long
as its ionization and thermal equilibration times are short compared to
the flow time, the gas must follow the equilibrium curve.  Because
the equilibrium curve lies along the vertical line $\Xi = \Xi_c$ until
the temperature exceeds $\sim 10^6$~K, the gas's ionization parameter
remains fixed at that value until the majority of its mass reaches
high temperatures.

    In fact, both the ionization and thermal balance timescales are indeed
considerably shorter than the flow time over a wide range of conditions.
The flow timescale
$t_{flow} \equiv r/v \sim 6 \times 10^{10} r_{\rm pc} (v/500$~\kms$)^{-1}$~s.
On the other hand, the thermal equilibration time
$t_{cool} \sim 2 \times 10^{7}$~s for our fiducial parameters
($\xi=100$, $T=1.5 \times 10^5$~K) when the gas is placed at 1~pc.
In Table~1 we list abundances and ionization equilibration timescales
$t_{ion}$ for commonly observed ions, calculated at ionization parameters
$\xi = 30$, 100, 300, and 1000 for gas illuminated at 1~pc by a central
source with an ionizing luminosity of $10^{44}~\rm erg~s^{-1}$.
For these parameters, the most abundant ions have
$t_{ion} \sim 10^{6 \pm 0.5}$~s.
Roughly speaking, $t_{cool,ion}/t_{flow} \propto r$,
so thermal and ionization equilibrium both become better approximations
when the gas is clumpier and closer to the nucleus, but even at
$r_{max}$ thermal and ionization equilibrium should prevail for
a substantial range of temperatures at $\Xi = \Xi_c$.

\begin{deluxetable}{lcccccccc}
\tabletypesize{\small}
%\rotate
\tablecaption{Equilibration times for selected ions}
\tablehead{
 & \multicolumn{2}{c}{$\xi = 30$}
 & \multicolumn{2}{c}{$\xi = 100$}
 & \multicolumn{2}{c}{$\xi = 300$}
 & \multicolumn{2}{c}{$\xi = 1000$}
\\
\colhead{Ion}
 & \colhead{A} & \colhead{$\tau_{\rm eq}$ (s)}
 & \colhead{A} & \colhead{$\tau_{\rm eq}$ (s)}
 & \colhead{A} & \colhead{$\tau_{\rm eq}$ (s)}
 & \colhead{A} & \colhead{$\tau_{\rm eq}$ (s)}
}
\small
\startdata
\ion{H}{1}    & $1.03 \times 10^{-7}$ & $2.79 \times 10^{1}$ & $1.10 \times 10^{-8}$ & $2.72 \times 10^{1}$ & $1.26 \times 10^{-9}$ & $2.75 \times 10^{1}$ & $1.86 \times 10^{-10}$ & $2.83 \times 10^{1}$  \\
\ion{He}{2}    & $2.42 \times 10^{-7}$ & $2.75 \times 10^{3}$ & $2.68 \times 10^{-8}$ & $2.72 \times 10^{3}$ & $3.17 \times 10^{-9}$ & $2.53 \times 10^{3}$ & $5.14 \times 10^{-10}$ & $2.62 \times 10^{3}$  \\
\ion{C}{3}    & $5.34 \times 10^{-9}$ & $1.46 \times 10^{3}$ & $2.54 \times 10^{-12}$ & $1.43 \times 10^{3}$ & $4.20 \times 10^{-16}$ & $1.32 \times 10^{3}$ & \nodata & \nodata  \\
\ion{C}{4}    & $2.37 \times 10^{-7}$ & $6.70 \times 10^{3}$ & $5.80 \times 10^{-10}$ & $6.66 \times 10^{3}$ & $1.43 \times 10^{-12}$ & $6.07 \times 10^{3}$ & $1.40 \times 10^{-14}$ & $6.24 \times 10^{3}$  \\
\ion{C}{5}    & $2.49 \times 10^{-5}$ & $6.92 \times 10^{4}$ & $5.62 \times 10^{-7}$ & $6.93 \times 10^{4}$ & $1.31 \times 10^{-8}$ & $6.96 \times 10^{4}$ & $5.02 \times 10^{-10}$ & $7.04 \times 10^{4}$  \\
\ion{C}{6}    & $4.10 \times 10^{-5}$ & $2.44 \times 10^{5}$ & $9.57 \times 10^{-6}$ & $2.45 \times 10^{5}$ & $1.38 \times 10^{-6}$ & $2.46 \times 10^{5}$ & $2.49 \times 10^{-7}$ & $2.49 \times 10^{5}$  \\
\ion{N}{5}    & $9.76 \times 10^{-7}$ & $1.75 \times 10^{4}$ & $4.94 \times 10^{-9}$ & $1.75 \times 10^{4}$ & $1.22 \times 10^{-11}$ & $1.66 \times 10^{4}$ & $8.58 \times 10^{-14}$ & $1.65 \times 10^{4}$  \\
\ion{N}{6}    & $2.20 \times 10^{-5}$ & $1.30 \times 10^{5}$ & $9.63 \times 10^{-7}$ & $1.30 \times 10^{5}$ & $2.29 \times 10^{-8}$ & $1.30 \times 10^{5}$ & $9.14 \times 10^{-10}$ & $1.32 \times 10^{5}$  \\
\ion{N}{7}    & $3.86 \times 10^{-4}$ & $4.21 \times 10^{5}$ & $1.65 \times 10^{-4}$ & $4.21 \times 10^{5}$ & $2.69 \times 10^{-5}$ & $4.23 \times 10^{5}$ & $4.98 \times 10^{-6}$ & $4.28 \times 10^{5}$  \\
\ion{O}{6}    & $3.87 \times 10^{-5}$ & $3.03 \times 10^{4}$ & $5.38 \times 10^{-7}$ & $3.04 \times 10^{4}$ & $1.62 \times 10^{-9}$ & $2.98 \times 10^{4}$ & $1.13 \times 10^{-11}$ & $2.94 \times 10^{4}$  \\
\ion{O}{7}    & $3.31 \times 10^{-4}$ & $2.24 \times 10^{5}$ & $3.81 \times 10^{-5}$ & $2.24 \times 10^{5}$ & $1.06 \times 10^{-6}$ & $2.25 \times 10^{5}$ & $4.25 \times 10^{-8}$ & $2.28 \times 10^{5}$  \\
\ion{O}{8}    & $4.87 \times 10^{-5}$ & $3.24 \times 10^{6}$ & $4.46 \times 10^{-5}$ & $6.78 \times 10^{5}$ & $9.18 \times 10^{-6}$ & $6.80 \times 10^{5}$ & $1.77 \times 10^{-6}$ & $6.89 \times 10^{5}$  \\
\ion{Ne}{8}    & $3.17 \times 10^{-5}$ & $7.88 \times 10^{4}$ & $2.10 \times 10^{-6}$ & $7.89 \times 10^{4}$ & $1.37 \times 10^{-8}$ & $7.91 \times 10^{4}$ & $1.18 \times 10^{-10}$ & $7.91 \times 10^{4}$  \\
\ion{Ne}{9}    & $5.74 \times 10^{-5}$ & $5.27 \times 10^{5}$ & $3.47 \times 10^{-5}$ & $5.28 \times 10^{5}$ & $1.83 \times 10^{-6}$ & $5.30 \times 10^{5}$ & $8.18 \times 10^{-8}$ & $5.37 \times 10^{5}$  \\
\ion{Ne}{10}    & $3.46 \times 10^{-6}$ & $1.49 \times 10^{6}$ & $1.86 \times 10^{-5}$ & $1.49 \times 10^{6}$ & $8.81 \times 10^{-6}$ & $1.50 \times 10^{6}$ & $2.01 \times 10^{-6}$ & $1.52 \times 10^{6}$  \\
\ion{Mg}{11}    & $6.33 \times 10^{-6}$ & $1.08 \times 10^{6}$ & $1.90 \times 10^{-5}$ & $1.08 \times 10^{6}$ & $3.43 \times 10^{-6}$ & $1.08 \times 10^{6}$ & $2.21 \times 10^{-7}$ & $1.10 \times 10^{6}$  \\
\ion{Mg}{12}    & $3.40 \times 10^{-7}$ & $2.84 \times 10^{6}$ & $8.61 \times 10^{-6}$ & $2.85 \times 10^{6}$ & $1.51 \times 10^{-5}$ & $2.86 \times 10^{6}$ & $4.91 \times 10^{-6}$ & $2.89 \times 10^{6}$  \\
\ion{Si}{13}    & $9.34 \times 10^{-7}$ & $1.98 \times 10^{6}$ & $1.01 \times 10^{-5}$ & $1.99 \times 10^{6}$ & $9.50 \times 10^{-6}$ & $1.99 \times 10^{6}$ & $1.09 \times 10^{-6}$ & $2.02 \times 10^{6}$  \\
\ion{Si}{14}    & $1.27 \times 10^{-8}$ & $4.91 \times 10^{6}$ & $1.03 \times 10^{-6}$ & $4.90 \times 10^{6}$ & $1.04 \times 10^{-5}$ & $4.92 \times 10^{6}$ & $5.89 \times 10^{-6}$ & $4.99 \times 10^{6}$  \\
\ion{S}{15}    & $4.75 \times 10^{-8}$ & $3.29 \times 10^{6}$ & $1.72 \times 10^{-6}$ & $3.29 \times 10^{6}$ & $9.44 \times 10^{-6}$ & $3.31 \times 10^{6}$ & $2.22 \times 10^{-6}$ & $3.35 \times 10^{6}$  \\
\ion{S}{16}    & $6.38 \times 10^{-11}$ & $2.97 \times 10^{12}$ & $1.82 \times 10^{-8}$ & $7.89 \times 10^{6}$ & $1.17 \times 10^{-6}$ & $7.92 \times 10^{6}$ & $1.36 \times 10^{-6}$ & $8.03 \times 10^{6}$  \\
\ion{Ar}{17}    & $3.32 \times 10^{-10}$ & $5.03 \times 10^{6}$ & $8.07 \times 10^{-8}$ & $5.04 \times 10^{6}$ & $1.51 \times 10^{-6}$ & $5.05 \times 10^{6}$ & $7.71 \times 10^{-7}$ & $5.12 \times 10^{6}$  \\
\ion{Ar}{18}    & $9.05 \times 10^{-13}$ & $1.05 \times 10^{6}$ & $1.85 \times 10^{-9}$ & $1.20 \times 10^{7}$ & $4.32 \times 10^{-7}$ & $1.20 \times 10^{7}$ & $1.08 \times 10^{-6}$ & $1.22 \times 10^{7}$  \\
\ion{Ca}{19}    & $2.79 \times 10^{-12}$ & $7.58 \times 10^{6}$ & $3.85 \times 10^{-9}$ & $7.58 \times 10^{6}$ & $4.06 \times 10^{-7}$ & $7.61 \times 10^{6}$ & $9.02 \times 10^{-7}$ & $7.71 \times 10^{6}$  \\
\ion{Ca}{20}    & $8.36 \times 10^{-14}$ & $2.32 \times 10^{3}$ & $9.17 \times 10^{-10}$ & $1.89 \times 10^{7}$ & $1.19 \times 10^{-6}$ & $1.75 \times 10^{7}$ & $1.29 \times 10^{-5}$ & $1.77 \times 10^{7}$  \\
\ion{Fe}{16}    & $6.35 \times 10^{-6}$ & $1.25 \times 10^{5}$ & $1.43 \times 10^{-6}$ & $1.25 \times 10^{5}$ & $1.45 \times 10^{-8}$ & $1.26 \times 10^{5}$ & $4.06 \times 10^{-13}$ & $1.27 \times 10^{5}$  \\
\ion{Fe}{17}    & $4.35 \times 10^{-6}$ & $1.78 \times 10^{5}$ & $9.40 \times 10^{-6}$ & $1.78 \times 10^{5}$ & $6.48 \times 10^{-7}$ & $1.79 \times 10^{5}$ & $7.61 \times 10^{-11}$ & $1.81 \times 10^{5}$  \\
\ion{Fe}{18}    & $6.71 \times 10^{-7}$ & $2.57 \times 10^{5}$ & $9.24 \times 10^{-6}$ & $2.58 \times 10^{5}$ & $4.33 \times 10^{-6}$ & $2.59 \times 10^{5}$ & $3.73 \times 10^{-9}$ & $2.62 \times 10^{5}$  \\
\ion{Fe}{19}    & $5.66 \times 10^{-8}$ & $2.72 \times 10^{5}$ & $3.81 \times 10^{-6}$ & $2.72 \times 10^{5}$ & $8.05 \times 10^{-6}$ & $2.73 \times 10^{5}$ & $4.84 \times 10^{-8}$ & $2.76 \times 10^{5}$  \\
\ion{Fe}{20}    & $5.27 \times 10^{-9}$ & $4.92 \times 10^{5}$ & $1.32 \times 10^{-6}$ & $4.93 \times 10^{5}$ & $1.08 \times 10^{-5}$ & $4.95 \times 10^{5}$ & $5.50 \times 10^{-7}$ & $5.01 \times 10^{5}$  \\
\ion{Fe}{21}    & $1.26 \times 10^{-10}$ & $3.86 \times 10^{5}$ & $1.06 \times 10^{-7}$ & $3.86 \times 10^{5}$ & $3.66 \times 10^{-6}$ & $3.88 \times 10^{5}$ & $1.13 \times 10^{-6}$ & $3.92 \times 10^{5}$  \\
\ion{Fe}{22}    & \nodata & \nodata & $4.05 \times 10^{-8}$ & $9.96 \times 10^{5}$ & $3.04 \times 10^{-6}$ & $1.00 \times 10^{6}$ & $4.83 \times 10^{-6}$ & $1.01 \times 10^{6}$  \\
\ion{Fe}{23}    & \nodata & \nodata & $4.62 \times 10^{-9}$ & $2.15 \times 10^{6}$ & $8.39 \times 10^{-7}$ & $2.16 \times 10^{6}$ & $6.88 \times 10^{-6}$ & $2.18 \times 10^{6}$  \\
\ion{Fe}{24}    & \nodata & \nodata & $1.08 \times 10^{-10}$ & $3.50 \times 10^{6}$ & $1.00 \times 10^{-7}$ & $3.51 \times 10^{6}$ & $5.09 \times 10^{-6}$ & $3.55 \times 10^{6}$  \\
\ion{Fe}{25}    & \nodata & \nodata & \nodata & \nodata & $6.15 \times 10^{-8}$ & $2.08 \times 10^{7}$ & $1.07 \times 10^{-5}$ & $2.11 \times 10^{7}$  \\
\ion{Fe}{26}    & \nodata & \nodata & \nodata & \nodata & $2.50 \times 10^{-9}$ & $4.68 \times 10^{7}$ & $2.20 \times 10^{-6}$ & $4.64 \times 10^{7}$  \\
\enddata
\tablecomments{Abundances relative to hydrogen (A) and
equilibration timescales ($\tau_{\rm eq}$)
for selected ions of the most astrophysically-abundant elements in
thermal balance at photoionization parameters of
$\xi = 30, 100, 300, 1000$ assuming
a distance of $r=1$~pc from a source of luminosity $10^{44}~\rm erg~s^{-1}$.
The computations were done using XSTAR Version 2.1.}
\end{deluxetable}

    At fixed radius and pressure, both $t_{cool}$ and $t_{ion}$ vary
along the vertical portion of the $T(\Xi)$ curve.
Roughly speaking, the thermal timescale grows $\propto T^2$ at constant
$\Xi$.  At the highest temperatures, however,
$t_{cool}$ grows more rapidly as line cooling is replaced by Compton cooling.
The Compton equilibration time is long enough, $\sim 10^{12} r_{\rm pc}^2
L_{44}^{-1}$~s, that fully reaching Comptonization equilibrium is
unlikely\footnote[1]{Note that the Compton time depends
on the total luminosity, not just the ionizing luminosity.}.  Moreover,
when $t_{cool} \sim t_{flow}$, adiabatic cooling can become important
(as in Krolik \& Begelman 1986 and Balsara \& Krolik 1993).  Thus, the
highest temperature on the equilibrium curve reachable by gas injected
cool is the temperature at which $t_{cool} \ltsim t_{flow}$, here
$T_{max} \simeq 4 \times 10^6 r_{\rm pc}^{-1/2}$~K (for distances
$\gtrsim 1$~pc; at smaller distances, $T_{max}$ approaches the Compton
temperature).  When the gas temperature is $\ltsim T_{max}$,
it is no longer constrained to lie on the equilibrium curve; due to
adiabatic cooling, it tends to move into the region of net heating.

    The ionization equilibration timescale (at fixed $\Xi$ but variable
$\xi$ or $T$) behaves in a somewhat more complicated fashion.  It
is determined by a combination of the ionizing flux and the local density.
The former dominates when the ionization rate for ion $j$ exceeds
the recombination rate from $j$ to $j-1$, the latter when the sense
of this comparison reverses.  When the most abundant ion of a given
element has charge $\geq j$, the former case generally applies.
Because many of the species most important for soft X-ray features
are H- and He-like and these elements are often mostly stripped, the
former case applies more often than the latter.

Although the flux at fixed photon energy is, of
course, entirely independent of $T$ at a given radius, as the gas
changes temperature, the photon energies relevant to its ionization
state change.  Both the photon fluxes and the cross sections can
be strong functions of energy.  For this reason, $t_{ion}$ also
generally rises with increasing temperature, but rather more slowly than
$t_{cool}$---for the dominant unstripped ion of each element, $t_{ion}$
increases by factors of $\sim 1$ -- 5 from the bottom to
the top of the vertical branch of the equilibrium curve.

   We conclude, then, that over a wide range of possible radii, the
thermal and ionization equilibration times are indeed short compared
to the flow time over most of the vertical section of the thermal
equilibrium curve.  After a flow time, it is possible for some gas
to reach high enough temperatures that it is no longer required to
be in thermal balance; however, it is not necessary that all the
gas reach this state.  Marginal thermal stability means that the temperature
of a given gas parcel can easily move either up or down on timescales
$\sim t_{cool}$; all
that is required is that its density change enough to compensate for
the temperature change.  Thus, we expect a range of temperatures to
be present in this region, and also a corresponding range of densities.

\subsection{Dynamics of the warm absorber}

    These considerations also constrain the absorber's dynamics.  The
absorber's equation of state is determined by how much of its material
is at low enough temperature to demand thermal balance.  When most
of the gas is in such a state, $\Xi$ is fixed at $\Xi_c$, and the
pressure must vary $\propto r^{-2}$.  The strength of the pressure
gradient relative to gravity is then given by
\begin{equation}
\label{qdef}
Q = {2 \over \eta_{out} \Xi_c} \left({L \over L_E}\right)^{1/4}
         \left({ \mu_e c^5 \over GM \sigma_T \sigma T_{c}^4}\right)^{1/4}
          \nonumber
  = 240 \left({L/L_E \over M_6}\right)^{1/4} \left({\Xi_c \over 15}\right)^{-1}
          \left(\eta_{out} T_{c3}\right)^{-1} .
\end{equation}
Here $\eta_{out}$ is the ratio of the mass loss rate to $L/c^2$, and
we suppose that $r$ is comparable to the inner edge of the torus,
where the radiation effective temperature is $T_c$.
The fiducial parameters in Equation~4 are $10^6 M_{\odot}$ for the
central mass $M$ and 1000~K for $T_c$ (the latter chosen with reference to the
results of Pier \& Voit 1995).

   For there to be outward acceleration, $Qu> (1 - \kappa L/L_E)$, where
$u$ is the speed in free-fall units and
$\kappa$ is the frequency-averaged opacity in Thomson units.  This condition
may be regarded as an upper bound on the mass loss rate in the sense
that $\eta_{out} \ltsim 240u(L/L_E)^{1/4}M_{6}^{-1/4}$ (assuming that
radiation forces do not
overcome gravity).  Strikingly, this bound roughly matches
the empirical estimate derived from eq.~\ref{massratio}. In terms
of the notation of that equation,
$\eta_{out} = \dot M_{out}/(\dot M_{acc} \eta_{acc})$.  Note, however,
when making this comparision, that formally $\dot M_{out}$ as estimated
in equation~\ref{massloss} reflects only the mass loss associated with
matter at a single value of $\xi$; if there is a distribution (as we
have now argued there should be), $\dot M_{out}$ is the sum of the
contributions from all the components (see also \S 3.2).

%\begin{equation}
%{\dot M \over L/c^2} = 170 {\Delta r \over r} \left({C \over 0.5}\right)
%         \left({\xi \over 2000}\right)^{-1} \left({L_{ion}/L \over 0.5}\right)
%         \left({v \over 500\hbox{~km~s$^{-1}$}}\right).
%\end{equation}
%For this estimate we normalize to $\xi = 2000$ and $\Delta r/r = 1$
%for the reasons given in \S 3.2.

   The observed spectra suggest that the radiation pressure contribution
is probably secondary.  The ratio between the amount of photon
momentum scattered by the wind and the total wind momentum is
\begin{equation}
f_{rad} = {\Delta L \over L}{\xi r/\Delta r\over 4\pi \mu_H c v^2}
        = 5 \times 10^{-3} \left({\Delta L / L \over 0.1}\right)
        \left({\xi \over 100}\right)(r/\Delta r)
        \left({v \over 500\hbox{~\kms}}\right)^{-2}
        \left({N(\xi=100) \over N_{tot}}\right),
\end{equation}
where the last ratio accounts for the range of ionization parameters
that may be present.   A scattered fraction $\Delta L/L \sim 0.1$ appears
to be consistent with what is seen \cite{Kaspi01}.  Unless $\Delta r/r$
is quite small, $f_{rad}$ is likely to be well less than unity.

     The pressure in the wind is $\propto \langle T \rangle \dot M
(u/r^{1/2})^{-1}r^{-2}$, where $\langle T \rangle $ is the mean temperature.
Fixed $\Xi$ then requires $\langle T \rangle \dot M (u/r^{1/2})^{-1}$
to be constant.  If $\dot M$ were constant (which is not strictly
required due to possible continuing evaporation or condensation), the
velocity (and $\langle T \rangle$) would increase only logarithmically
with $r$.  However, because the gas dynamics depend on its
pressure, not its temperature, there is no requirement for the
temperature to rise smoothly and homogeneously so long as $\Xi = \Xi_c$.
Temperature fluctuations with wavelengths shorter than $r$ and long
enough that thermal conduction does not damp them can be easily created.
We would expect, therefore, that, although there may be an overall
sense of temperature evolution, there can be substantial local
inhomogeneity.
Eventually, the approximation made at
the beginning that $\Xi = \Xi_c$ must break down.  As
Compton equilibrium is neared, the equilibrium $\Xi$ begins to
increase and the requirement of tracking the equilibrium curve is
weakened as $t_{cool}$ approaches and then exceeds $t_{flow}$.

\section{Comparison with observations}

\subsection{Inferred ionization parameters and temperatures}

    As we have just seen, this model predicts that the distribution
of $\Xi$ in this gas should be very narrow, but the temperature
distribution is likely to be much wider.
In a fashion strikingly consistent with these predictions, Kaspi
et al. (2001) found it impossible to adequately fit their data with
only a single component.  Moreover, their two inferred components
lie squarely on the vertical section of the equilibrium curve, one
near $8 \times 10^4$~K, the other at $\simeq 6 \times 10^5$~K.  The
similarity of all the line profiles suggests that these two
components are subject to more or less the same forces, and are
therefore not very distant from each other.  Because they have
the same value of $\Xi$, if they lie at similar distances from
the central source, their pressures must also be the same.  Only
if there is global pressure-confinement (i.e., a volume-filling
background medium) would one expect these pressures to match.
    
    Our model predicts temperatures across a somewhat wider
range, from $\simeq 3 \times 10^4$~K up to $\sim 10^6$~K.  It
may be difficult to detect the hottest gas, however.
As one can see in Table~1, only for elements
with atomic number $\geq 12$ (Mg and above) do more than 10\% of
the atoms retain at least one electron when $\xi = 1000$ ($T \simeq 9
\times 10^5$~K); only a small increase in temperature beyond that
point suffices to eliminate almost all the unstripped ions except
Fe.  In the band covered by the Chandra gratings, only
two lines in gas with $\xi \simeq 1000$ (\ion{Fe}{22}~10.98,11.77)
can be expected to have line-center optical depths greater than
unity when the velocity span is 500~km~s$^{-1}$ and the column
density is $10^{22}$~cm$^{-2}$ (although another half dozen lines have
opacities a factor of two smaller).  Moreover, because the H-like
stages of Mg, Si, S, Ar begin to be abundant when $T > 5 \times 10^5$~K,
it can be hard to distinguish $10^6$~K gas from gas at half that
temperature.

     UV features, on the other hand, grow in opacity at the lowest
temperatures.  For example, Table~1 shows that the fractional
abundances\footnote[2]{XSTAR uses the elemental abundances of
Grevesse et al. 1996.}
of \ion{C}{4} and \ion{O}{6} relative to H rise from $5.8 \times 10^{-10}$
and $5.4 \times 10^{-7}$, respectively, at $\xi = 100$ ($1.5 \times 10^5$~K)
to $2.4 \times 10^{-7}$ and $3.9 \times 10^{-5}$ at $\xi = 30$
($4.7 \times 10^4$~K).
Any significant amount of gas at the low end of the temperature range is then
likely to generate UV absorption features.   These are, in fact,
often observed.  In order to produce the line strengths seen,
not much is required: for typical cases
(Crenshaw et al. 1999; Kriss et al. 2000; Kaiser et al. 2001),
$\ltsim 10^{19}$~H~cm$^{-2}$ suffices if the gas
temperature is $5 \times 10^4$~K.

\subsection{Clumping}

   If there is nothing to restrain it, there is no reason why this
gas should not expand to fill the volume available to it.  In fact,
if the absorber gas were not volume-filling, one would have to
ask what gas occupied the remainder of the region, and that
gas would be subject to the same thermodynamic constraints as
the gas under consideration.  Within this volume, however, there
are no obvious constraints favoring any particular temperature from
$\simeq 3 \times 10^4$~K up to the maximum temperature achievable within
a flow time, a few $\times 10^6$~K.

   We do not know how to predict the distribution of column density
with temperature $dN/dT$.  Once it is given, though, we can find
the distribution of volume with temperature:  A differential
column density $dN$ having volume density $n$ occupies a radial shell
with thickness $d\Delta r = dN/n$.  In terms of the pressure $p$,
$d\Delta r = (kT/p)dN$, so that the distribution of fractional
volume with temperature $d(\Delta r/r)/dT \propto T dN/dT$.
The normalization of this relation is fixed by the constraint
$\int \, dT \, d(\Delta r/r)/dT = 1$.
That is, for equal column density and fixed pressure,
higher temperature regions occupy larger volumes.

   Given a particular distribution $dN/dT$, it is possible to
predict the distance of the entire system from the central
continuum source.  Suppose, for example, that $dN/d\log T$
is constant from $T_{min} = 3 \times 10^4$~K up to
$T_{max} = 10^6$~K.  Because
$\xi \propto \Xi/T$, from Equation~1 it then follows that
the distance at which this material exists is
\begin{eqnarray}
\label{r_pred}
 r_{pred} \simeq {L_{ion} \log (T_{max}/T_{min}) \over
              4\pi c \Xi_c kT_{max} N_{tot}}
   \simeq 4 L_{ion,44}
   \left({N_{tot} \over 3.5 \times 10^{22}\hbox{~cm$^{-2}$}}\right)^{-1}
   \left({\Xi_c \over 15}\right)^{-1}
                \left({T_{max} \over 1 \times 10^6\hbox{~K}}\right)^{-1}
                \hbox{~pc},
\end{eqnarray}
where $N_{tot}$ is the total column density, approximately a
factor of 3.5 greater than the column density per $\log T$.
Unfortunately, this estimate is rather model-dependent because
we have no way of determining the true $dN/dT$.  In particular,
to the degree that the higher temperatures have larger column
densities, $r_{pred}$ would decrease.  However, this example
does emphasize the likelihood that gas at intermediate $T$
occupies only a fraction of the volume.

\subsection{Variability}

    For considerations of bulk dynamics, ionization balance is a good
approximation when $t_{ion} < t_{flow}$.  However, because we
can observe changes on timescales much shorter than $t_{flow}$,
it is possible for us to see continuum fluctuations
driving short-lived changes in the ionization balance.  In fact,
because the observed continuum flux and the absorption features
``ride the light-cone" together, we can even follow fluctuations
on timescales much shorter than the light-crossing time.  The
column densities we measure reflect the history
of the continuum flux over the past $t_{ion}$; the temperature
inferred from column densities is the product of heating and
cooling over the past $t_{cool}$.

     The relation between observed absorption column densities
and the continuum provides an excellent diagnostic of the
actual distance between the continuum source and the gas.
At a distance of 1~pc, Table~1 shows $t_{ion} \sim 10^{6 \pm 0.5}$~s
for those species most visible in the soft X-ray band.
However, since $t_{ion} \propto r^2$, the magnitude of $t_{ion}$
depends sensitively on the column density in the hottest
(and nearly transparent) component.  For example, if the
guess illustrated in equation~\ref{r_pred} is correct,
the true distance would be $\simeq 4$
times larger, making the ionization timescales $\simeq 16$
times longer, or $10^{7.2 \pm 0.5}$~s for the species creating
soft X-ray features.  UV features should respond somewhat more
quickly.  Although it may be very difficult to monitor AGN
X-ray lightcurves with sampling density sufficient to quantitatively
match continuum fluctuations with ionization fluctuations, it
should at least be possible to estimate upper bounds on ionization
response times.  These bounds can then provide constraints on the
distance to this material, and, in this model, $dN/dT$.

     As of now, there has been little in the way of systematic
monitoring of warm absorbers.  The clearest report in the literature
of very rapid changes in an absorbing column is an apparent sharp increase
in the \ion{O}{8} K-edge optical depth in MCG--6-30-15 that
occurred in a span of $\sim 10^4$~s during an ASCA observation
(Reynolds et al. 1995; Otani et al. 1996).  Within the context of
our model, this time could
be brought into agreement with the observed time if there were
a column density $\simeq 2 \times 10^{23}$~H~cm$^{-2}$ in gas at
$T \gtrsim 10^6$~K.  However, it is also possible that
this feature may have been confused
with \ion{O}{8} K$\alpha$ emission broadened by relativistic
motions in the accretion disk \cite{BR01}.  The fact
that when the \ion{O}{8} column density in MCG~-6-30-15
increased, the measured \ion{O}{7} column density was constant
may support this alternate interpretation.  A somewhat similar
event occurred in NGC~4051, but with the roles of \ion{O}{7}
and \ion{O}{8} reversed---the \ion{O}{7} column density changed
while the \ion{O}{8} column density was constant \cite{G96}.
In this case, because the luminosity of the nucleus is so small
($\sim 10^{42}$~erg~s$^{-1}$), the timescale---$\sim 10^4$~s---places
only a weak upper bound on the distance, $r \ltsim 0.2$~pc.  This
is, of course, easily consistent with estimates like equation~\ref{r_pred}.

    Evidence in the UV to date is similarly fragmentary.  Kriss
et al. (1997) reported Lyman edge optical depth changes in NGC~4151,
and Espey et al. (1998) saw \ion{C}{3} column density changes in the
same object that
anti-correlated with the continuum flux on timescales of days.
Unfortunately, the X-ray absorption in NGC~4151 is so complex that the
warm absorber column density cannot be estimated well enough for
us to predict the ionization timescale.

\subsection{Emission lines and the covering factor}

   The peculiar geometry of flow through the opening in toroidal
obscuration means that we cannot see the entire emission portion of
P~Cygni profiles created by absorption lines.  In part this is
because at many radii the warm ionized gas simply does not fill
all solid angle (most of it is occupied by the dusty, molecular
obscuring gas); in part this is because the obscuring matter
partially blocks our view of the far side of the flow.  In line with this
prediction, Kaspi et al. (2001) use the P~Cygni features they
discovered in the spectrum of NGC~3783 to estimate covering fractions of
$\simeq 0.3$ and $\simeq 0.5$ for their high- and low-ionization
components, respectively.

    Unless $r \ll r_{max}$, variations in the column densities of the
emission components on the timescale of continuum fluctuations
(days to weeks) should be averaged out
because the light-crossing time is much longer (several
years).  Consequently, the covering fraction inferred on the basis
of a single ``snap-shot" may not be a fair measure of the true
covering fraction.

\section{Summary}

     A wide variety of evidence now points toward locating the gas
that produces ionized absorption features in the soft X-ray spectra
of AGN in a warm wind that lies a few parsecs from the nucleus in
Seyfert galaxies.  Both absorption and emission lines are seen with
widths of $\sim 500$~\kms, in many cases with the telltale ``P Cygni"
velocity shifts.  The inferred ionization parameters and column
densities indicate maximum distances $\sim 30$~pc.

     The specific ionization states indicated by the observed lines
suggest that there are a number of components with different
values of $\xi$, the density form of the ionization parameter.
However, these all have the same value ($\Xi = \Xi_c$) of the pressure form of
the ionization parameter.  Above $\Xi_c$, low-temperature
equilibria do not exist, and the temperature rises from $\simeq 3
\times 10^4$~K to $\simeq 10^6$~K with $\Xi$ very nearly constant at $\Xi_c$.
If these components coexist at the same distance from the central
source, they share the same pressure, suggesting a volume-filling
configuration by which the pressure is regulated.  If the volume-filling
phase is the highest attainable temperature on the marginally stable
branch of the thermal equilibrium curve, the actual distance could
be substantially smaller than the maximum; a factor of ten smaller
would not be surprising.

     Because the thermal equilibria at $\Xi = \Xi_c$ are marginally
stable to fixed-pressure thermal perturbations, material can
move easily from one temperature to another.  The only constraint
is that accessible temperatures must have thermal equilibration times
short compared to the flow time.  Consequently, we expect
that multi-temperature models will in general be required in order to
explain all the observed features.

    Given that a very similar wind in a very similar
location has been previously inferred from the scattering of nuclear
light in obscured Seyfert galaxies, it makes sense to suppose that
the two are essentially the same structure.  However, absorption
features in type~1 Seyfert galaxy X-ray spectra provide much more
sensitive diagnostics of its state than does the polarized reflection
of the continuum that can be observed in type~2 Seyfert galaxies.
In particular, the time-dependent response of the absorbing column densities
as the X-ray continuum level changes should provide a very sensitive
test of these ideas.

\acknowledgments

    We would like to thank Mike Crenshaw for organizing the workshop
``Mass Outflow in Active Galaxies: New Perspectives", whose discussions
provoked some of these thoughts.  We are also indebted to Tim Kallman for
creating and maintaining the new version of XSTAR.  This work was partially
supported by NASA Grant NAG5-9187 to JHK.

\end{document}